\documentclass[aps,prl,twocolumn,showpacs]{revtex4}

\usepackage{graphicx}

\begin{document}

\title{Strictly In-Plane Vortex-like Excitations in Pseudogap Region of Underdoped $La_{2-x}Sr_xCuO_4$ Single Crystals}

\author{H. H. Wen$^1$, Z. Y. Liu$^1$, Z. A. Xu$^2$, Z. Y. Weng$^3$, F. Zhou$^1$, Z. X. Zhao$^1$}

\affiliation{$^1$ National Laboratory for Superconductivity,
Institute of Physics, Chinese Academy of Sciences, P.~O.~Box 603,
Beijing 100080, P.~R.~China}

\affiliation{$^2$ Physics Department, Zhejiang University,
Zhejiang, P.~R.~China}

\affiliation{$^3$ Center for Advanced Study, TsingHua University,
Beijing 100084, P.~R.~China}

\date{\today}

\begin{abstract}
Nernst effect has been measured for an underdoped and an optimally
doped $La_{2-x}Sr_xCuO_4$ single crystal with the magnetic field
applied along different directions. For both samples, when $H \|$
c, a significant Nernst voltage appears above $T_c$ and keeps
measurable up to a high temperature ( 100 K ) as reported by Xu et
al.[ Nature 406, 486(2000) ]. However, when $H \|$ a-b plane the
Nernst signal drops below the noise level quickly above $T_c$ for
the underdoped sample. Moreover, the in-plane Nernst data with
field at different angles shows a nice scaling behavior with the
c-axis component of the field. All these give strong indication
for strictly in-plane vortex-like excitations in pseudo-gap
region.
\end{abstract}


\maketitle

One of the puzzles in high temperature superconductors is the
origin of a pseudogap above $T_c$ in underdoped region.
Experimentally it is clearly
seen\cite{Loeser,DingHong,Loram,Optical,NMR} that some electronic
density of states ( DOS ) near the Fermi surface is depleted below
the pseudogap temperature $T^*$ which is much higher than $T_c$.
In order to understand the physics behind the pseudogap, many
models have been proposed, such as resonating valence bond ( RVB
)\cite{RVB} theory and its bosonic version called phase string
theory\cite{StringPhase}, spin fluctuation\cite{SpinFluc},
preformed Cooper pairs\cite{Emery}, charge
stripes\cite{EmeryKivelson}, d-density wave
(DDW)\cite{Chakravarty,Affleck}, etc. Among many of them, the
pseudogap state has been considered as the precursor to
superconducting state. In this precursor state, composite bosons,
such as holons or pairs of electrons have been formed before the
long range phase coherence ( or Bose-Einstein like condensation )
is established. Measurements on the high frequency complex
conductivity illustrate that a short-life phase coherence can
persist up to about 30 K above $T_c$ in the short time scale and
the ultra-fast electrical response can be described by the
thermally generated topological defects ( free vortices ) in that
phase\cite{Corson}. Similar conclusions are drawn in understanding
the coherent peak of ARPES\cite{FengDL} and the ( $\pi,\pi$ )
magnetic resonance peak at 41 meV in inelastic neutron
scattering\cite{DaiPC} in the pseudogap region. By doing thermal
measurement, the Princeton group found that a significant Nernst
signal\cite{Xu} appears in the pseudogap region with H$\|$ c. This
may be understood by the phase-slip due to the thermal drifting of
vortex-like excitations. However, it is still unknown whether this
excitation should be strictly in-plane and how the direction of
the external field influences this excitation.

Moreover, given the formation of Cooper pairs above $T_c$, it is
desirable to know whether the superconducting condensation occurs
simultaneously with the coherent motion of Cooper pairs along
c-axis. Presumably there are two major pictures. The first one is
that a finite phase stiffness in the superconducting state is
established by the binding of the topologically generated
vortex-antivortex pairs\cite{Emery,Corson,Deutscher}. In this
frame $T_c$ is corresponding to the Kosterlitz-Thouless transition
and is not necessarily related to the c-axis coherent motion of
quasi-particles or Cooper pairs. Another one is that the
condensation occurs by establishing the coherent motion of Cooper
pairs between layers, i.e., the interlayer tunnelling model
(ILT)\cite{ILT}. The ILT model assumes that the driving force for
the pairing and thus the condensation ( in the frame of BCS
picture ) is the reduction of kinetic energy when holes recover
the c-axis coherent motion at $T_c$, thus the condensation energy
and the Josephson coupling energy should be equal. In this paper
we present data of Nernst effect near and above $T_c$ in
underdoped and optimally doped $La_{2-x}Sr_xCuO$ single crystals
with the magnetic field aligned in different directions. Our
results prove a strictly in-plane vortex-like excitation above
$T_c$ and the superconducting condensation occurs simultaneously
with the coherent motion of Cooper pairs along c-axis in
underdoped region.

\begin{figure}
\includegraphics[scale=1]{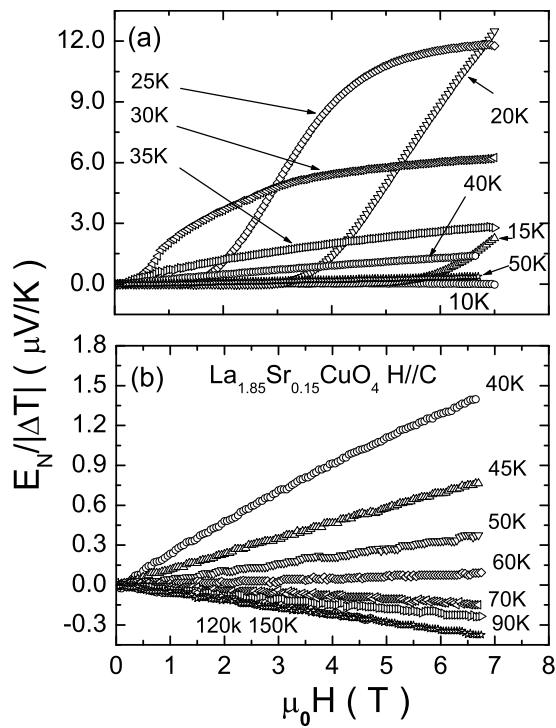}
\caption{ Nernst signals for the optimally doped sample when
$H\|c$ at temperatures ranging from (a)10 K to 50 K; and (b) 40 K
to 150 K. A background from the normal holes ( electrons ) is
shown by the data of 120 K and 150 K. Both the magnitude and the
temperature region for the Nernst signal is very similar to that
reported by Xu. et al.\cite{Xu}} \label{fig1}
\end{figure}

Single crystals measured in this work were prepared by the
travelling solvent floating-zone technique. In this Letter we
present the measurements for two typical single crystals, one
underdoped ( x = 0.09, $T_c = 30 K$) and another one optimally
doped ( x = 0.15 $T_c = 38.5 K$) as characterized by AC
susceptibility and DC magnetization. The qualities of our samples
have also been characterized by x-ray diffraction patterns showing
only (00l) peaks, rocking curve showing a full width at the half
maximum ( FWHM ) of ( 008 ) peak being narrower than 0.10
$^\circ$, the RBS channelling minimum yield being only
$\chi_m$=0.038, and resistive measurement showing a rather narrow
transition width $\Delta T_c \leq $ 1.5 K.

For measuring the Nernst effect we adopted the
one-heater-two-thermometer technique. A heating power of 1-4 mW is
applied to one end of the single crystal and two tiny thermometers
with distance of 1.5 mm are attached to the sample for measuring
the temperatures along the heat flow ( longitudinal ) direction.
The Nernst voltage is measured through two contacts on two
opposite planes at the symmetric positions ( shown by inset in
Fig.2 ). Very small contacting resistance ( $\leq 0.1 \Omega$ )
has been achieved by using silver paste. Both samples are shaped
into a bar structure with dimensions of $3-4 mm ( length ) \times
1 mm ( width ) \times 0.5 mm ( thickness )$. All measurements are
based on an Oxford cryogenic system ( Maglab-12 ) with temperature
stability better than 0.01 K and magnetic field up to 12 Tesla.
During the measurement for Nernst signal the magnetic field is
swept between 7 to -7 Tesla and the Nernst signal $V_N$ is
obtained by subtracting the positive field value with the negative
one in order to remove the Faraday signal during the field sweep
and the possible thermal electric power due to asymmetric electric
contacts. The Nernst voltage is measured by a Keithley
182-Nanovoltmeter with a resolution of about 8 nV in present case.
In this paper we show the Nernst signal $E_N/|\Delta T|$, where
$E_N = V_N/d$ with $d$ the distance between the two contacts for
Nernst voltage, $\Delta T$ is the temperature gradient along the
heat flow direction.

In Fig.1(a) we present the Nernst voltage $E_N$ vs. external
magnetic field ( $H\|c$ ) for the optimally doped sample at
temperatures ranging from 10 K to 50 K. It is clear that in low
temperature region with a vortex solid state, $E_N$ is too small
to be measurable below 7 tesla ( see the curve at 10 K ). At 15 K,
when H is above 5 T, the Nernst signal appears and rises showing a
pronounced vortex motion. The onset field for the appearance of
$E_N$ is corresponding very well to the irreversibility line
measured on the same sample in the magnetization measurement. The
signal $E_N$ first rises ( see, e.g., 20 K and 25 K ) and then
drops gradually when the temperature is further increased ( see,
e.g., 30 K, 35 K, 40 K and higher ). For a conventional
superconductor, below $T_c$, the Nernst signal should first
increase, reach a maximum and then decrease and finally drop to
zero near $H_{c2}$. Wang et al.\cite{WangYYPRL2002} measured the
Nernst effect in high field region and find that the upper
critical field $H_{c2}(T)$ extends to a much higher temperature
and field region, especially for underdoped samples. From our data
one can also see that at 50 K, $E_N(H)$ is still in the uprising
branch which agrees with the basic idea about a high $H_{c2}$ of
Wang et al.\cite{WangYYPRL2002} The data for temperatures higher
than $T_c$ are shown in Fig.1(b). Just as reported by Xu et
al.\cite{Xu}, a significant Nernst signal is observed far above
$T_c$. From our data here $E_N$ is measurable up to about 100 K
where it approaches the background signal evidenced by almost
identical values measured at 120 K and 150 K. Similar phenomenon
occurs for the underdoped sample ( x = 0.09 ) which reveals a
sizable Nernst signal up to about 120 K. As forementioned this
significant signal was attributed\cite{Xu} to the vortex-like
excitations in the pseudogap region. Both the general magnitude
and the temperature region for the measurable Nernst voltage when
$H\| c$ on our crystals are close to that found by Xu et
al.\cite{Xu} indicating an unambiguous repeatable experimental
discovery.

\begin{figure}
\includegraphics[scale=0.8]{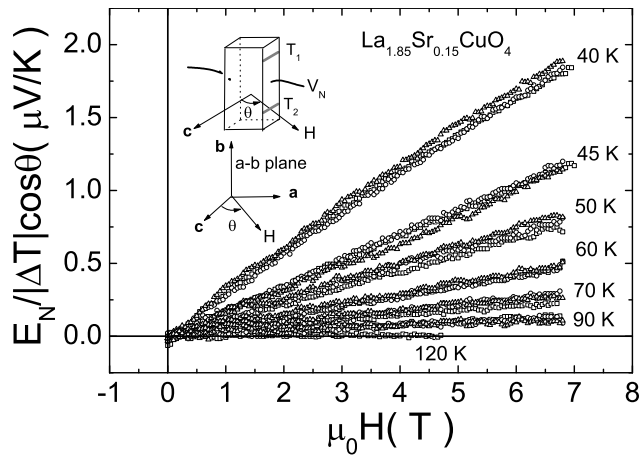}
\caption{  Scaling of the Nernst signal versus the angle between
the direction of the magnetic field and c-axis. The three angles
are $\theta$ = 0$^\circ$ ( squares ), 25$^\circ$ ( circles ), and
50$^\circ$ ( triangles ), respectively. It is clear that the
in-plane Nernst signal can be well scaled with $E_N \propto H cos
\theta $. All data are subtracted with the background at 150 K.
The inset shows the configuration of our measurement, where a, b
and c correspond to the crystalline axes.}\label{fig:fig2re}
\end{figure}

In order to know whether this normal state Nernst signal is due to
an in-plane vortex-like excitation, we now turn the direction of
the magnetic field away from c-axis. The angle between the
magnetic field ( always in the a-c plane ) and c-axis is $\theta$.
The configuration of the measurement is shown as an inset in
Fig.2. If $E_N$ is contributed by strictly in-plane vortex-like
excitations, it should be scalable when $\theta$ is changed. The
data collected at temperatures ranging from 40 K to 90 K and
$\theta = 0^\circ, 25^\circ, 50^\circ$ are shown in Fig.2. Because
from 40 K to 90 K it is found that $E_N \propto H$, in order to
expose the scaling quality of the data at different temperatures,
here we show $E_N/cos\theta$ vs H, which should be the same as
$E_N$ vs H$\times cos \theta$. It is easy to see that the three
set of data corresponding to three angles can be well scaled
together at each temperature. Although the Nernst voltage can be
written in a general way $V_N= \alpha \Delta THcos\theta$, it is
not naturally true that $V_N$ is strictly proportional to
$Hcos\theta$ since the pre-factor $\alpha$ here is also $\theta$
dependent. One can expect that $V_N$ is strictly proportional to
$Hcos\theta$ only when the system is isotropic or strictly 2D in
nature, since otherwise the vortex core size, shape and the motion
viscosity coefficient will depend on $\theta$, leading to a angle
dependence of $\alpha$. This can be proved by the failure of
scaling the data below $T_c$ in the flux flow region where the
moving pancakes are connected by Josephson vortices. Nice scaling
obtained here above $T_c$ shows that the in-plane Nernst signal
depends only on the c-axis component of the magnetic field. This
shows a strong evidence that the normal state Nernst signal is
induced by strictly in-plane vortex-like excitations.

Now we turn to the situation when $H\|ab$ plane and $E_N$ is
measured along c-axis. For the underdoped sample ( x = 0.09 ), it
is found that the Nernst signal when $H\|c$ behaves very similar
like that shown in Fig.1(a) and (b) for the optimally doped
sample, i.e., a significant signal $E_N$ appears up to a high
temperature ( ~ 120 K ). For brevity, the data is not shown here.
However, when $H\|ab$ plane, the situation becomes very different.
Firstly, as shown in Fig.3, it is clear that below $T_c$, the
signal $E_N$ becomes much smaller than that when $H\| c$. This can
be understood with different vortex structure and dynamics in
these two different cases. At $H\|c$, the vortex system are
composed by pancake vortices which are connected by Josephson
vortices. The dissipation induced by flux flow due to thermal
gradient is mainly contributed by vortex pancakes. Since these
pancake vortices have much smaller volume as well as smaller
collective pinning energy $U_c$, the motion is relatively easier
comparing to the case with $H\|ab$. When $H\|ab$ plane, the
Josephson vortex is driven to move by the thermal drifting force.
\begin{figure}
\includegraphics[scale=0.8]{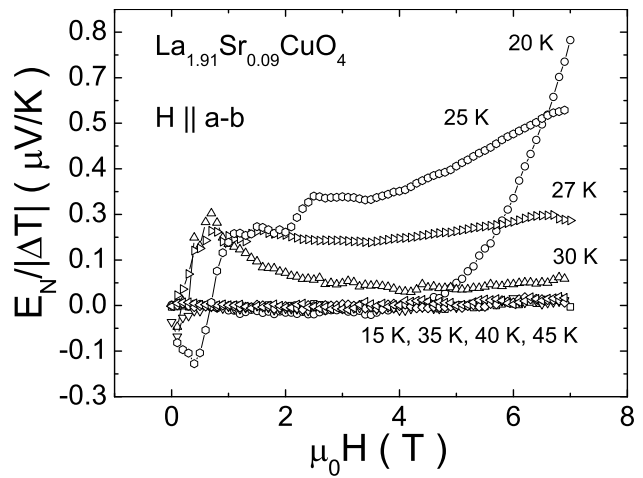}
\caption{ Nernst signal for the underdoped sample when $H\| a-b$
plane. The signal in the superconducting state is much smaller
than that when $H\| c $. Remarkably, the Nernst signal vanishes
quickly above $T_c = 30 K$.} \label{fig3}
\end{figure}
Each Josephson vortex lies in between two neighbor Cu-O planes and
threads through a large part or even the whole sample. Due to the
squeezing effect by the supercurrent in neighboring planes, the
vortex size along the moving direction is $\lambda_c /\epsilon$,
where $\lambda_c$ is the penetration depth along c-axis,
$\epsilon$ is the anisotropy of the mass matrix elements along ab
plane and c-axis ( $\epsilon \approx$ 1/20 for LSCO ). Because of
this special structure, the moving speed of these Josephson
vortices is relatively lower leading to a smaller Nernst signal.
Besides, in Fig.3, one can see another remarkable point that the
signal $E_N$ drops abruptly to zero ( the background $10^{-8}$ V )
when the temperature is just above $T_c$. The vanishing of $E_N$
above $T_c$ is certainly not due to the relatively smaller signal
in the superconducting state at $H\|ab$ since a clear step rather
than a gradual manner appears near $T_c$.
\begin{figure}
\includegraphics[scale=0.8]{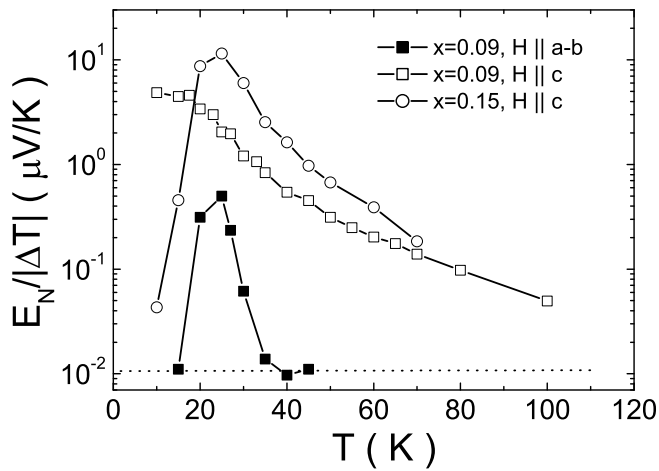}
\caption{ Temperature dependence of Nernst signal at 6 Tesla for
the underdoped sample ( $T_c = 30 K$, squares ) and the optimally
doped sample ( $T_c = 38. 5 K$, circles ). The dashed line
represents the noise level of our measurement. } \label{fig4}
\end{figure}
In order to have a closer inspection, in Fig.4 we present the
Nernst signal at 6 T for $H\|c$ and $H\|ab$ at different
temperatures for two samples\cite{OptimalDoped}. One can see that
$V_N$ can be measurable up to about 100 K when $H\|c$, but it
drops below the noise level at about 35 K which is only 5 K from
$T_c$ when $H\|ab$. If the sizable signal $V_N$ measured below
$T_c$ reflects the motion of the Josephson vortices when $H\| ab$
plane, the vanishing $V_N$ above $T_c$ may suggest that
superconducting condensation is established almost simultaneously
with the Josephson coupling along c-axis. Since Josephson coupling
is fulfilled by the coherent motion of Cooper pairs along c-axis,
it is also tempting to conclude that the superconducting
condensation occurs simultaneously with the coherent motion of
Cooper pairs along c-axis. Similar conclusion was drawn in earlier
infrared reflectivity (IR) measurements by Tamasaku et
al.\cite{Uchida} who observed a sharp reflectivity edge which was
attributed to Josephson plasma resonance. However, beyond the
earlier IR data, the present results show clear evidence for
in-plane vortex-like excitation\cite{WengZY} which implies the
existence of preformed pairing far above $T_c$. In this scenario
the ILT model may be modified in the following way: the
superconducting condensation ( in underdoped region ) occurs by
lowering the kinetic energy of the Cooper pairs which are
inhibited to move along c-axis above $T_c$. Thus the total
condensation energy has two contributions: reduction of kinetic
energy of Cooper pairs below $T_c$ and the energy reduction due to
pairing above $T_c$. This picture naturally explains why the
Josephson coupling energy as observed in some systems, such as,
Tl-2201, is smaller than the total condensation energy determined
from specific heat measurement\cite{VanderMarel,Moler}. A similar
picture is proposed by Chakravarty et al.\cite{Chakravarty} who
argued that the condensation energy cannot be precisely defined
since the normal state is not a Fermi liquid. Therefore this
picture calls for further quantitative theoretical consideration.

In conclusion, beside a strong Nernst signal observed for
underdoped and optimally doped samples above $T_c$ when $H\|c$,
for the first time a general scaling for the Nernst signal vs.
c-axis component of the magnetic field is obtained. Furthermore,
for the underdoped sample the Nernst signal vanishes abruptly just
above $T_c$ when $H\| ab$ plane showing that the superconducting
condensation is established simultaneously with the Josephson
coupling ( or the coherent motion of Cooper pairs ) along c-axis.
Both observations strongly suggest the existence of strictly
in-plane vortex-like excitations.

\section{Acknowledgments}

This work is supported by the National Science Foundation of China
(NSFC 19825111, 10274097), the Ministry of Science and Technology
of China ( project: NKBRSF-G1999064602 ), the Knowledge Innovation
Project of Chinese Academy of Sciences. We are grateful for
fruitful discussions with T. Xiang at Institute of Physics (
Chinese Academy of Science ), Yayu Wang at Princeton University.




\begin{thebibliography}{00}
\bibitem{OptimalDoped}It is worth noting that the Nernst signal
measured when $H\|ab$ plane for the optimally doped sample can
also keep a high value to a very high temperature ( even above 200
K ) which is far above $T^*\approx$ 100 K. The signal lies much
higher than the value measured when $H\|c$ in all temperature
regions above $ T_c$. Detailed analyses reveal that this strong
anomolous Nernst signal is due to the uncanceled part of the
normal electrons. This interesting discovery together with more
data will be presented in a separate paper.
\end{thebibliography}
\end{document}